\documentclass[3p,times,twocolumn,procedia]{elsarticle}

\usepackage{ecrc}

\volume{}

\firstpage{1}

\journalname{Nuclear Physics B Proceedings Supplement}

\runauth{A. Denig}

\jid{nuphbp}

\jnltitlelogo{Nuclear Physics B Proceedings Supplement}

\CopyrightLine{2014}{Published by Elsevier Ltd.}

\usepackage{amssymb}

\usepackage[figuresright]{rotating}

\begin{document}

\begin{frontmatter}



\dochead{}

\title{Measurements of the Hadronic Cross Section and of Meson Transition Form Factors at BESIII for an improved Standard Model Prediction of $(g-2)_\mu$}


\author{Achim Denig\\
for the BESIII collaboration}

\address{Institute for Nuclear Physics and PRISMA Cluster of Excellence, Johannes Gutenberg University Mainz, Johann-Joachim-Becher-Weg 45, D-55099 Mainz, Germany}

\begin{abstract}
The BESIII experiment at the Beijing tau-charm factory BEPCII recently has embarked on a series of form factor 
measurements with the goal to improve the hadronic vacuum polarization as well as the
hadronic light-by-light contributions to $(g-2)_\mu$. The status of the flagship measurements and 
preliminary results are presented. 
\end{abstract}

\begin{keyword}
anomalous magnetic moment of the muon \sep hadronic cross section \sep meson transition form factors \sep 
precision test of the Standard Model \ gamma gamma physics \ R measurements


\end{keyword}

\end{frontmatter}


\section{The Hadronic Contribution to the Anomalous Magnetic Moment of the Muon}
\label{introduction}

The anomalous magnetic moment of the muon $a_\mu \equiv (g-2)_\mu/2$~\cite{jeg} has been measured a few years ago by the BNL-E821 experiment with an accuracy of better than 1 part in $10^{6}$: 
$a_\mu^{\rm exp} = ( 11 659 208.9 \pm 5.4_{\rm stat} \pm 3.3_{\rm syst} ) \cdot 10^{-10}$ ~\cite{Bennett:2006fi}.
Within the Standard Model (SM) $(g-2)_\mu$ can be calculated with slightly better accuracy.
This allows for a unique test of the SM of particle physics. It is well known, that the SM calculation does not only
receive contributions from QED, but also from weak and strong interactions and probably also from Beyond SM (BSM) interactions. Presently, the experimental and SM
values of $(g-2)_\mu$ differ by more than 3 standard deviations~\cite{Davier:2010nc}~\cite{Hagiwara:2011af}, which triggered many speculations
whether this might be an indication of a missing BSM contribution. 
\begin{figure}[h]
\centerline{\includegraphics[width=0.5\textwidth]{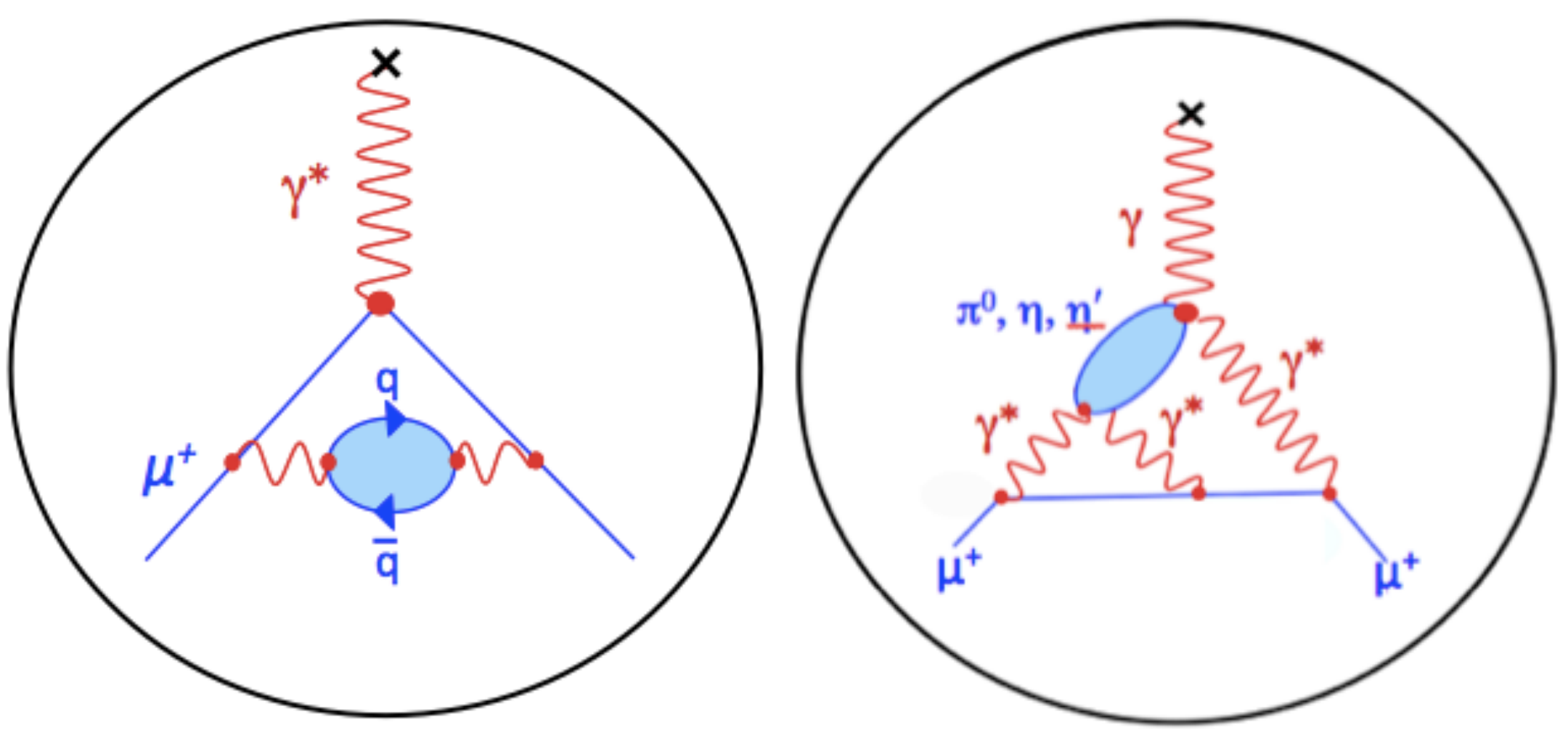}}
\caption{Hadronic contributions to $(g-2)_\mu$: the hadronic vacuum polarization (left) and 
the hadronic Light-by-Light contribution (right).}
\label{Fig:1}
\end{figure}
New and improved measurements of 
$(g-2)_\mu$ at FNAL˜\cite{fnal} and JPARC˜\cite{jparc} are upcoming and it is hence important for the final interpretation of these new results to improve also
the SM prediction of $(g-2)_\mu$. Here, the most important issue is 
the strong interactions contribution.  The total contribution is only 60ppm, but
it completely dominates the SM uncertainty. 
The hadronic contribution is split into two main parts, namely
the Hadronic Vacuum Polarization (HVP), seen in the left Feynman diagram of Fig.~\ref{Fig:1}, and the
Hadronic Light-by-Light Scattering (HLbL), seen in the right Feynman diagram of Fig.~\ref{Fig:1}, contributions. 
\\
It is well known that measurements of the {\bf hadronic cross section in $\mathbf{e^+e^-}$ annihilation} at low energies can be used to compute the HVP part via a dispersion integral. Using this approach, the leading order HVP 
contribution is found to be $a_\mu^{\rm HVP}=(692.3)_{\pm4.2} \cdot 10^{-10}$, therefore being the leading
source of uncertainty in the SM prediction.
 The HLbL contribution up to now
is calculated using hadronic models. In most compilations of $(g-2)_\mu$ the result of Prades, de Rafael, and Vainshtein ~\cite{prades} is used: $a_\mu^{\rm HLbL}=(10.5)_{\pm2.6} \cdot 10^{-10}$. Very recently, it was
proposed~\cite{bern}~\cite{mainz} that also in the case of the HLbL contribution, experimental data on {\bf meson transition form factors} 
could be used, which would
for the first time allow for a purely phenomenological calculation of a major part of HLbL. 
\\
In the following chapters we will discuss the impact of the BESIII experiment at the tau-charm factory BEPCII 
in improving the HVP as well as HLbL contributions to $(g-2)_\mu$. 
A description of the detector can be found 
elsewhere~\cite{bes3}. For the results shown in this paper a data set of 2.9 fb$^{-1}$ taken a center-of-mass energy of 3.770 GeV is used.
A larger data set of approximately 5 fb$^{-1}$ is available above 4 GeV and is already 
used for various analyses.

\section{Measurement of the Hadronic Cross Section}
\label{R}

Presently, the exclusive final states with 2 pions, 3 pions, and 4 pions are analyzed at BESIII. We 
present in this paper only results for the two-pion final state $\sigma(e^+e^- \to \pi^+\pi^-)$, which contributes to more
than 70\% to the total HVP correction. We briefly comment also on the perspectives for R measurements at BESIII
at the end of this chapter.
\\
Unfortunately, the BaBar measurement of $\sigma(e^+e^- \to \pi^+\pi^-)$~\cite{babar}, which has a claimed
systematic accuracy of 0.5\%, shows quite some deviation from measurements of KLOE~\cite{kloe}, which
itself claims a 0.8\% accuracy for the most precise of its data sets. The deviation is in the order of 3\% on the $\rho$ peak and increases towards
higher energies. Precision data points from Novosibirsk~\cite{cmd2}~\cite{snd} have larger statistical and systematic uncertainties and hence can confirm neither the BaBar nor the KLOE results. 
\\
As in the case of KLOE and BaBar, the initial state radiation (ISR) method is used by BESIII, in which events are analyzed, in which one of the beam electrons/positrons has
emitted a high-energetic photon (intial state radiation, ISR)~\cite{isr}. Depending on the energy of the
ISR photon, the hadronic mass in the final state is reduced and the hadronic cross section can be extracted 
for all masses below the actual center-of-mass energy of the collider, $\sqrt{s}=$3.770 GeV. 
In order to be competitive with the most precise data sets for the two-pion channel, an accuracy at the
1\% level is required. 

\begin{figure}[hb]
\centerline{\includegraphics[width=0.5\textwidth]{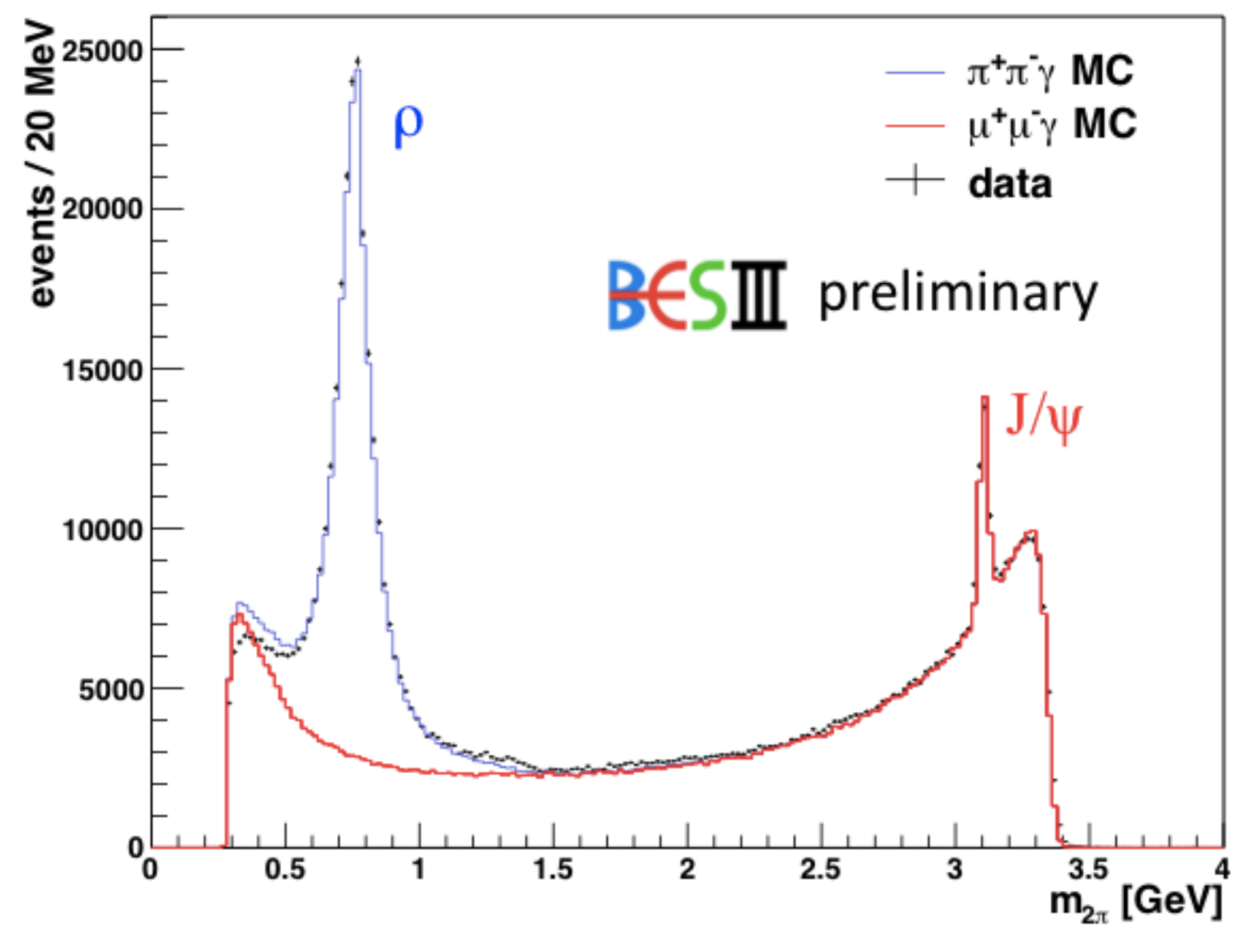}}
\caption{Yield of $e^+e^- \to \pi^+\pi^-\gamma$ events before the dedicated background subtraction.}
\label{Fig:2a}
\end{figure}

\subsection{ISR analysis of ${e^+e^- \to \pi^+\pi^-}$}
\begin{figure}[hb]
\centerline{\includegraphics[width=0.55\textwidth]{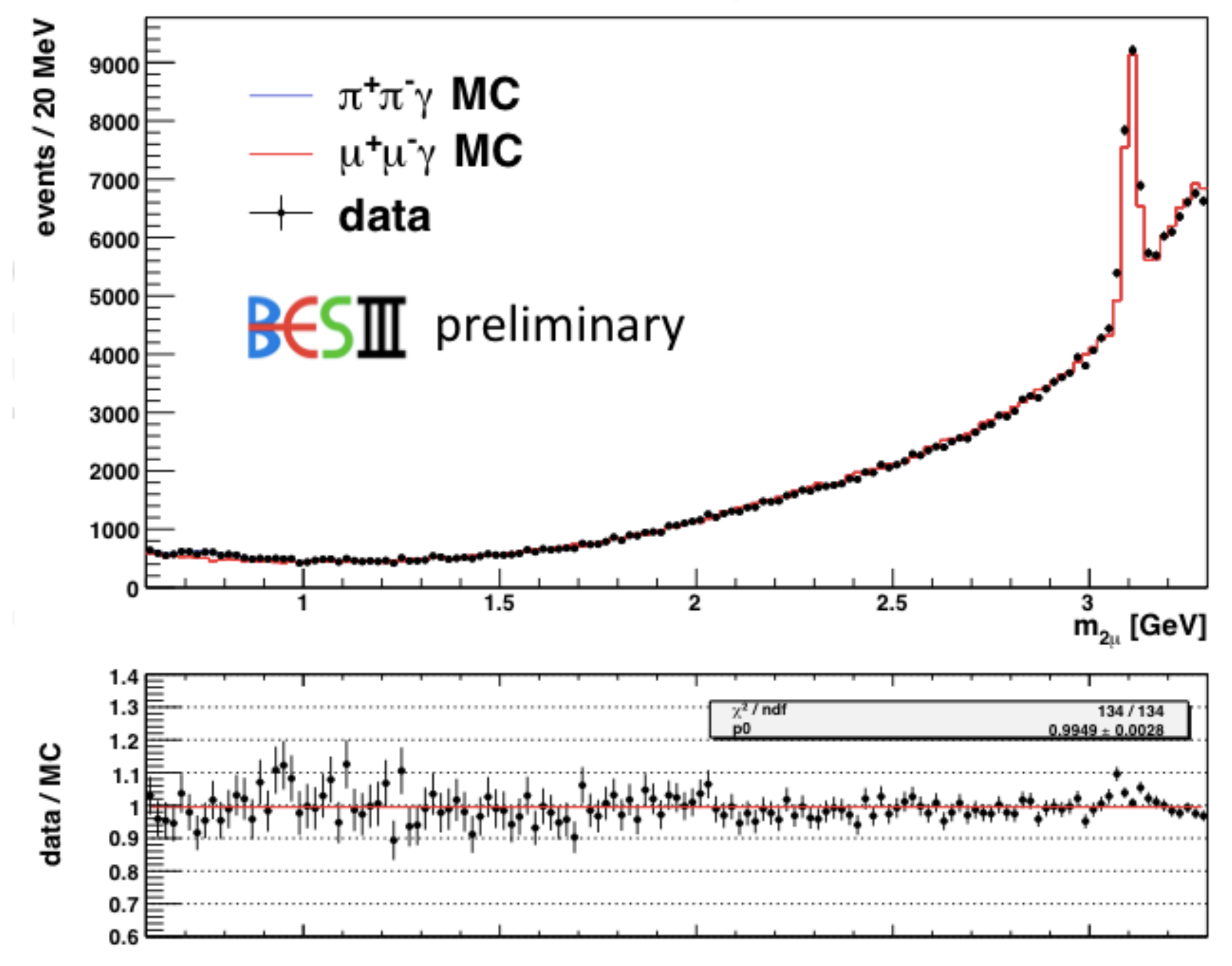}}
\caption{Above: Absolute comparison of data and the PHOKHARA MC~\cite{mcmumu} for the process $e^+e^- \to \mu^+\mu^-\gamma$. Below: Relative difference. An accuracy of better than 1\% is achieved.}
\label{Fig:2}
\end{figure}

\begin{figure}[h]
\centerline{\includegraphics[width=0.55\textwidth]{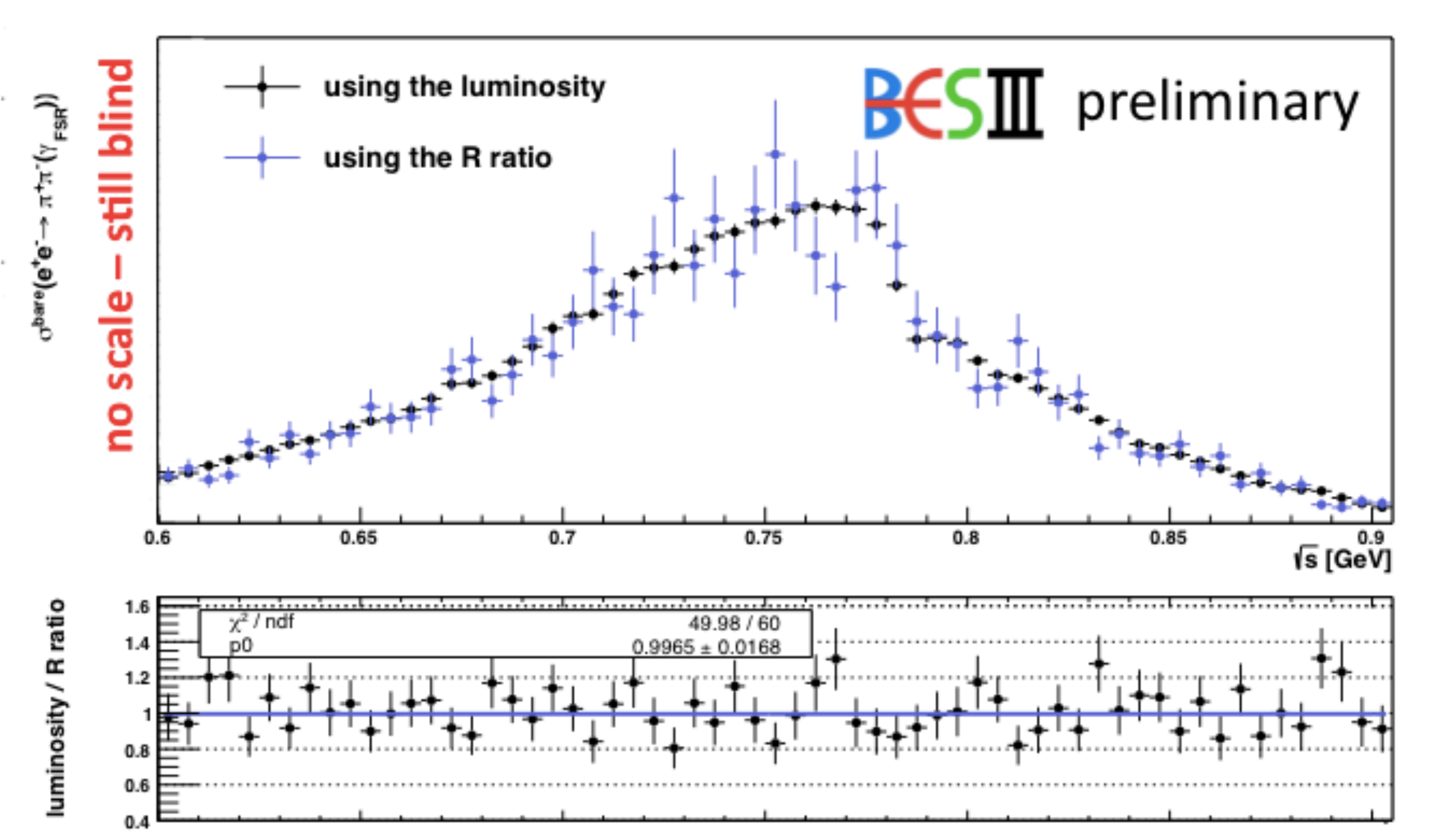}}
\caption{Upper panel: Extracted cross section $\sigma(e^+e^- \to \pi^+\pi^-)$ obtained from a normalizations to the integrated luminosity (black points) and to $e^+e^- \to \mu^+\mu^-\gamma$ events (blue points). The vertical 
scale is still hidden to avoid a bias from previous measurements.
Lower panel: Relative difference between the two normalization methods.
}\label{Fig:4}
\end{figure}

In the past, ISR measurements have been performed with and without the
detection of the ISR photon (tagged and untagged analyses). Only in the latter case small angle ISR photons
are included in the selection, which in turn yields a very high event rate. At BESIII the kinematics allows for an untagged
analysis above approximately 1 GeV of hadronic masses. In the work presented here, we will concentrate on the
$\rho$(770) peak region and we will discuss cross section measurements between 600 and
900 MeV. Hence, we perform a tagged analysis and the ISR photon is detected in the electromagnetic
calorimeter. 
\\
After an initial event selection, in which in addition to the ISR photon the presence of two charged tracks of opposite charge from the interaction point are required, a 4C kinematic fit is 
performed in the signal hypothesis $e^+e^- \to \pi^+\pi^-\gamma$. 
Fig.~\ref{Fig:2a} shows the event yield after this initial selection as a function of the two-pion
invariant mass $m_{2\pi}$. A clear $\rho$ peak is visible, which is due to signal events. An enormous background from
$e^+e^- \to \mu^+\mu^-\gamma$ events as well as data-MC discrepancies are also seen. 
\\
In order to remove the muon background, an artificial neural network has been trained to determine the
probability for a track to be identified as a pion or muon. Also a track-based correction between
signal MC and data has been determined using independent control samples. 
An important cross check of this muon-pion
separation is a data-MC comparison for muon events. Unlike in the pion case, for the muon case a direct
comparison with QED is possible. The result is shown in Fig.~\ref{Fig:2}. 
We find an excellent agreement of better
than 1\% between data and the PHOKHARA MC generator ~\cite{mcmumu}, which was interfaced with the detector simulation package. 
While the systematic uncertainty of data is slightly larger than 1\%, the PHOKHARA accuracy is better than 0.5\%. 
\\
For the extraction of the cross section $\sigma_{\pi\pi} \equiv \sigma(e^+e^- \to \pi^+\pi^-)$ from the ISR data, the following
master formula is used:
\begin{equation}
\sigma_{\pi\pi}=\frac{N_{\pi\pi\gamma}/\epsilon_{\rm exp}}{L_{\rm int} \cdot H_{\rm rad} \cdot \delta_{\rm vac}
\cdot(1+\delta_{\rm FSR})}.
\end{equation}
$N_{\pi\pi\gamma}$ denotes here the number of selected signal events, $\epsilon_{\rm exp}$ the signal
efficiency, $L_{\rm int}$ the integrated luminosity~\cite{lumi}, and $H_{\rm rad}$ the
theoretical radiator function, which has been obtained from PHOKHARA, where the ISR process is calculated 
with NLO precision. The efficiency $\epsilon_{\rm exp}$ has also been obtained from PHOKHARA, but
additional data-MC corrections due to imperfections of the detector simulation are taken into account. Those corrections have
 been measured by means of control samples. Two additional corrections
are needed, namely the vacuum polarization correction $\delta_{\rm vac}$ to obtain the {\it bare} cross section, and a correction from final state radiation (FSR) effects. An unfolding of the mass spectrum has been
performed using the SVD approach~\cite{svd}. 
\\
A preliminary result of the two-pion cross section
is shown in Fig.~\ref{Fig:4} (upper panel, black data points). At this point, the result is still blind to avoid a possible bias from
previous measurements and therefore the scale on the vertical axis is not displayed. 
The blue data points show the
cross section obtained by using a different normalization, namely a normalization to $e^+e^- \to \mu^+\mu^-\gamma$ events. In this case, it is well known that several important contributions in Equation 1, and systematic effects related to them, cancel out. This concerns for instance the radiator function $H_{\rm rad}$, or the
integrated luminosity $L_{\rm int}$. Unfortunately, the muon statistics is limiting the accuracy for this method,
such that the approach shown in Equation 1 gives more accurate results for the presently analyzed data set. A comparison of the two
normalization methods is shown in the lower panel of Fig.~\ref{Fig:4}. The two results differ by $(0.35\pm1.68)\%$, where the large error is due to the low muon statistics. For the final publication the normalization to the
integrated luminosity will therefore be used and the muon normalization serves only as a rough cross check.

\subsection{R scan in the energy range 2.0 - 4.6 GeV}

The BEPCII collider 
allows for a direct energy scan 
in the center-of-mass energy range between 2.0 to 4.6 GeV . Recently, the range above 3.8 GeV has 
been covered in a high-statistics scan with fine energy binning. This allows to study
the resonance spectrum in the charmonium region with unprecedented accuracy.
For the near future a high statistics R scan below 3 GeV is scheduled with the goal
to measure the timelike form factors of nucleons~\cite{denig}
as well as of hyperons. Besides the cross section measurements, the goal is to extract the 
$G_E/G_M$ ratios from the angular distributions. In the case of the $\Lambda$ form factors, for the first time
the relative phase between $G_E$ and $G_M$ can be extracted. 
The combination of the high energy and low energy  scan together with a few additional data points 
between 3.0 and 3.8 GeV will allow to measure the inclusive hadronic cross section normalized to the di-muon
cross section, 
$R_{\rm incl} \equiv \sigma_{\rm incl}(e^+e^- \to {\rm hadrons})/\sigma(e^+e^-\to\mu^+\mu^-)$, with unprecedented accuracy
and to improve upon the so-far world's best measurements from BES and BESII~\cite{besR}. The overall goal
is to measure $R_{\rm incl}$ with a systematic uncertainty of approximately 3\%, which corresponds
to an improvement of a factor 2. The statistical uncertainty per scan point, which was in the
case of the old BES measurement between 3 -- 5 \%, will be reduced to below 1\%. 
This will not only have implications for the HVP contribution of 
$(g-2)_\mu$, but more importantly also for the determination of the hadronic corrections to $\alpha_{\rm em}(M_Z^2)$, which is an important precision observable for electroweak fits to the SM.

\section{Measurement of Meson Transition Form Factors}
\label{gg}

As mentioned above, the new dispersive approaches~\cite{bern}~\cite{mainz} to the HLbL contribution of $(g-2)_\mu$ motivate
precise measurements of meson transition form factors. Those form factors are already now very relevant as they allow to validate some of the hadronic models used in the standard calculations~\cite{nyffeler}. 
 More specifically, form factors with one or two non-vanishing virtualities of the photons are needed,
$\gamma^* \gamma^{(*)} \to P$, where P=$\pi^0, \eta, \eta^\prime$. Also two-body states, like $\pi\pi$ and
$\pi\eta$ are of highest interest. In order to make an impact for HLbL, data at low momentum transfer is
needed. Indeed, it has been shown that new and precise data from the B factory experiments BaBar and 
Belle~\cite{bfactory} on the $\pi^0$ transition form factor at high momentum transfer does not have a 
major impact. 
\\
For the time being, the BESIII collaboration has embarked on measurements
of the form factors of the lightest pseudoscalar states as well as the $\pi^+\pi^-$ state using 2.9 fb$^{-1}$ of data
taken at $\sqrt{s}=$ 3.770 GeV. We present in this paper the status of the $\pi^0$ form factor, which is measured in the spacelike domain exploiting events, in which the hadronic state has been produced in the fusion of the two photons emitted from the beam particles ($\gamma\gamma$ scattering).  
\\
The single-tag technique is used, in which one of the beam leptons is scattered at large polar angles and hence is
detected in the fiducial volume of the BESIII detector. 
By measuring
the momentum of the hadronic final state (by exclusivery detecting all its decay products) and by calculating the 
missing momentum, it can be guaranteed that the second beam lepton is almost unscattered. In such a way, it is controlled that the virtuality of the second photon is quasi-real. The process under study is
therefore $e^+e^- \to e^+e^- {\rm hadron(s)}$, where either the electron or positron in the final state is detected.
The
form factor as a function of one virtuality, $F(Q^2)$, is measured in this configuration, where $Q^2$ is the virtuality of the
photon associated to the lepton scattered at large angles.  
\\
\begin{figure}[h]
\centerline{\includegraphics[width=0.5\textwidth]{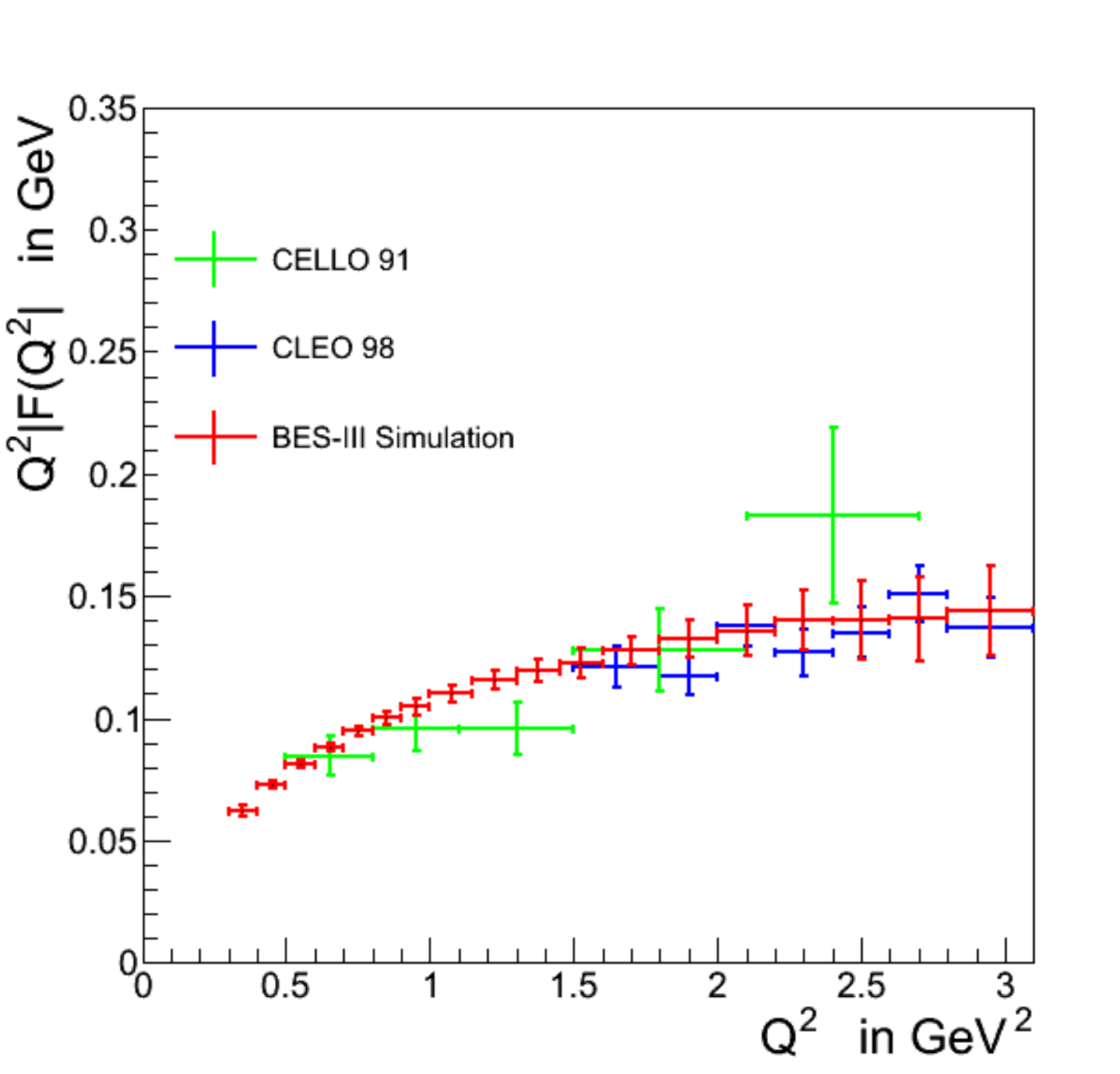}}
\caption{Existing data sets from CLEO~\cite{cleo} and CELLO~\cite{cello} for the pion transition form factor in comparison to the accuracy which will be achieved by BESIII, see text for explanation.}\label{Fig:5}
\end{figure}
\\
For the form factor measurement of the $\pi^0$, $F_\pi(Q^2)$, events with
the following signature are selected: $e^+e^- \to e^+e^- \pi^0 \to e^+e^- \gamma\gamma$. One
of the electrons or positrons is not detected, as explained above. A huge background from Bhabha
scattering processes and other QED processes has to be rejected. By looking at the two-photon 
invariant mass distribution, the two photons from the $\pi^0$ decay are selected. 
The missing momentum is calculated taking into account the known four momentum of the initial state, the
momentum of the scattered electron/positron and of the two photons from the $\pi^0$ decay. 
Clear peaks at cos$\theta = \pm 1$ are seen, where $\theta$ is the angle of the missing momentum. These peaks
correspond to the unscattered beam positron or electron of signal events.
\\
The number of signal events as a function of $Q^2$ is finally determined by looking at the yield of $\pi^0$ events in the
$\gamma\gamma$ invariant mass distribution. For the simulation of the signal MC the EKHARA event generator ~\cite{ekhara} is used. 
\\
Fig.~\ref{Fig:5} shows existing data from CLEO ~\cite{cleo} and CELLO ~\cite{cello} on $F_\pi(Q^2)$ in the
$Q^2$ range below 3 GeV$^2$ in comparison to the expected accuracy from BESIII. 
As the result is not yet publically available,
the central values of the BESIII spectrum in Fig.~\ref{Fig:5} correspond to the dual-octet model used in EKHARA. The size of the error bars, however, demonstrate what realistically has been achieved in the event analysis. 
We see a significant improvement in statistics and quality in the relevant $Q^2$ range below 2 GeV$^2$. We plan to publish the measurement of the pion transition form factor within the coming months.

\section{Conclusions}

A programme of ISR, $R_{\rm incl}$, and $\gamma\gamma$ measurements has been launched at the BESIII experiment. This programme is motivated by an improved determination of the HVP and HLbL contributions to $(g-2)_\mu$.
For the ISR hadronic cross section measurements, data taken at $\sqrt{s}=$ 3.770 GeV is analyzed.
An ISR measurement of the flagship analysis $\sigma(e^+e^- \to \pi^+\pi^-)$ is close to completion with an
expected accuracy of approximately 1\%. A normalization to the integrated luminosity will be used
for this measurement. Presently, the statistics of radiative di-muon events is still limiting a
normalization to muons, as previously done by KLOE and BaBar. With future runs foreseen at BESIII at
$\sqrt{s}=$ 3.770 GeV, or by including the existing high-energy runs, the statistics of $e^+e^- \to \mu^+\mu^-\gamma$ events will largely improve. 
A large data set above $\sqrt{s}=$ 3.770 GeV is available already now and is indeed extremely useful for the 
$\gamma\gamma$ programme as the cross section for these events increases with $\sqrt{s}$.
\\
An improved SM prediction of $(g-2)_\mu$ is of utmost importance in view of 
the new FNAL and JPARC measurements. The BESIII results will contribute to this worldwide effort as discussed in a recent whitepaper~\cite{whitepaper}.

\label{conclusion}





\bibliographystyle{elsarticle-num}
\bibliography{<your-bib-database>}



\end{document}